\documentclass{article}
\usepackage{spconf,amsmath,amssymb,graphicx,graphicx}
\usepackage{bm}
\usepackage{algorithm}
\usepackage{algorithmic}
\usepackage{booktabs}
\usepackage{multirow}
\usepackage{subcaption}
\usepackage{hyperref}
\usepackage{xcolor}

\def\L{{\mathcal L}}
\def\F{{\mathcal F}}
\def\D{{\mathcal D}}
\def\C{{\mathcal C}}
\def\E{{\mathcal E}}

\title{LUMEN: Low-light Unified Multi-stage Enhancement Network using depth-guided flash, clustering, and attention-based Transformers}

\name{Bibhabasu Debnath, Sahana Ray, and Sanjay Ghosh, Senior Member, IEEE}
\address{Department of Electrical Engineering, IIT Kharagpur, WB 721302, India}

\begin{document}
\maketitle

\begin{abstract}
Low-light image enhancement remains a challenging problem due to severe noise, color distortion, contrast degradation, and loss of structural details under insufficient illumination. Existing methods typically apply uniform enhancement without considering the depth-dependent nature of light attenuation and sensor noise in real-world scenes. To address this limitation, we propose LUMEN, a multi-stage enhancement framework that integrates virtual flash simulation with transformer-based feature fusion. The proposed framework first estimates scene depth from low-light inputs using a dedicated encoder–decoder network, after which a soft clustering module partitions pixels into depth-aware regions, enabling depth-dependent flash simulation. The simulated flash features, together with depth representations, are fused with image features through efficient attention-based fusion blocks to enhance global context while preserving fine details. A composite loss function combining reconstruction, perceptual, structural, color, edge, and depth consistency objectives ensures both visual fidelity and perceptual quality. Extensive experiments on LOL-v1 and LOL-v2 benchmarks demonstrate that LUMEN achieves state-of-the-art performance and produces visually natural results compared with several state-of-the-art methods. 
\end{abstract}

\begin{keywords} 
Low-light enhancement, depth estimation, flash simulation, transformer fusion, image restoration
\end{keywords}

\section{Introduction}
\label{sec:intro}

Images captured under poor lighting conditions present significant challenges for both human perception and computer vision applications, particularly in object detection and scene understanding tasks. When optical sensors operate with insufficient illumination or encounter rapidly varying external lighting, the resulting images exhibit multiple degradation artifacts: amplified sensor noise, severely reduced brightness levels, diminished contrast ratios, unnatural color shifts due to limited dynamic range, and substantial loss of fine textural details \cite{guo2016lime}. As a result, low-light image enhancement (LLIE) has emerged as a vital research area with broad applications in surveillance, autonomous navigation and medical imaging.

Traditional approaches rely on hand-crafted priors and classical image processing. Histogram equalization methods \cite{chen2003minimum, ibrahim2007brightness} redistribute pixel intensities to expand dynamic range. Retinex-based decomposition techniques \cite{ghosh2019fast1, ghosh2019fast, land1967lightness} model images as the product of reflectance and illumination maps, aiming to normalize lighting while preserving scene structure. LIME \cite{guo2016lime} employs an optimization-based framework. 

Deep learning methods have dramatically advanced LLIE performance through data-driven feature learning \cite{debnath2025low}. Retinex-inspired networks like KinD \cite{zhang2019kindling} employs dedicated subnetworks for separate reflectance and illumination restoration, while Retinex-Net \cite{wei2018deep} introduces a Decom-Net followed by illumination-adjusted enhancement; more recent advances like URetinex-Net++ \cite{wu2025interpretable} integrate implicit regularization priors and cross-stage feature fusion. Unsupervised methods like Zero-DCE \cite{guo2020zero} predicts high-order curve parameters for non-linear intensity mapping, while EnlightenGAN \cite{jiang2021enlightengan} employs lightweight adversarial training.
Despite these advances, most methods apply uniform processing without considering the inherent 3D geometric structure of natural scenes. Low-light degradation exhibits strong depth dependence: nearby objects receive higher illumination while distant regions experience exponential attenuation along with atmospheric scattering and sensor noise. Real flash photography addresses this through distance-based falloff, but introduces uneven lighting and appearance of harsh shadows.

In this work, we propose LUMEN, a depth-guided flash-simulated low-light image enhancement framework that explicitly models scene geometry to enable spatially adaptive illumination correction. The framework first estimates scene depth from low-light inputs using a dedicated encoder–decoder network supervised by pretrained MiDaS guidance. A differentiable soft K-means clustering module then partitions the scene into depth-aware regions, allowing physically motivated virtual flash simulation that selectively enhances dark distant areas while preserving naturally illuminated foreground regions. The generated flash representation is encoded and fused with depth and image features through Efficient Fusion Blocks incorporating multi-head transformer attention, enabling global contextual reasoning and multi-scale feature interaction.

The remainder of this paper is organized as follows. Section~\ref{sec:method} presents the proposed LUMEN framework, including the depth estimation network, differentiable depth-based clustering, flash simulation module, and transformer-based enhancement architecture. Section~\ref{sec:loss} describes the composite training objective and individual loss components used for optimization. Section~\ref{sec:exp} provides implementation details, datasets, evaluation metrics, and quantitative comparisons with state-of-the-art methods. Finally, Section~\ref{sec:conclusion} concludes the paper and outlines future research directions.

\section{Proposed Method}
\label{sec:method}

\begin{figure*}
\centering
\includegraphics[width=0.95\textwidth]{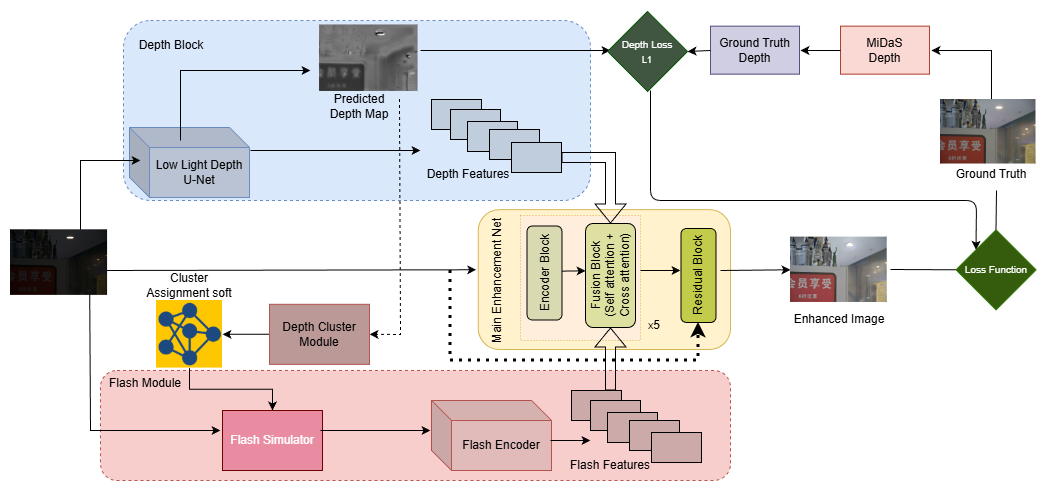}
\caption{The input low-light image is processed by a Low Light Depth U-Net (blue) to estimate a depth map, supervised by MiDaS during training. This depth map conditions a Depth Cluster Module (red) to segment image regions, enabling an adaptive Flash Simulator to generate a detail-enhanced guide image. A Flash Encoder extracts multi-scale features from this guide. The Main Enhancement Net (yellow) integrates these hierarchical depth features (spatial structure) and Flash features (texture details) via Efficient Fusion Blocks employing self- and cross-attention mechanisms at each encoder level. A global residual connection yields the final enhanced output.}
\label{fig:UNet}
\end{figure*}

Given a low-light input image $I_{low} \in \mathbb{R}^{H \times W \times 3}$, our goal is to produce an enhanced image $I_{enh}$.
The complete framework is illustrated in Fig.~\ref{fig:UNet}.

\subsection{Low-Light Depth Estimation Network}
\label{ssec:depth}

For accurate depth estimation, we employ an encoder–decoder network $\D$ with five hierarchical encoder levels using channel sizes $[C, 2C, 4C, 8C, 16C]$, where $C=64$. Each level applies max-pooling followed by double convolution blocks with batch normalization and LeakyReLU ($\alpha=0.2$).
The decoder mirrors the encoder with bilinear upsampling, skip connections, and double convolution refinement. A final $1 \times 1$ convolution with sigmoid activation produces the normalized depth map $D_{pred} \in [0,1]^{H \times W}$:
\begin{equation}
    D_{pred}, \{F^{(l)}_d\}_{l=1}^{5} = \D(I_{low}),
\end{equation}
where $\{F^{(l)}_d\}_{l=1}^{5}$ denotes intermediate encoder features at each level $l$.

\subsection{Depth-Based Clustering Module}
\label{ssec:cluster}

A differentiable depth-based clustering module $\C$ groups pixels according to their depth values using a soft K-means formulation ($K=8$).
Let $\{\mu_k\}_{k=1}^{K}$ denote learnable cluster centers initialized uniformly in $[0,1]$. For each pixel location $(i,j)$, we compute the soft cluster assignment as:
\begin{equation}
    A_{k}(i,j) = \frac{\exp(-|D_{pred}(i,j) - \mu_k|/\tau)}{\sum_{k'=1}^{K}\exp(-|D_{pred}(i,j) - \mu_{k'}|/\tau)},
\end{equation}
where $\tau=0.1$ is a temperature parameter controlling assignment sharpness. The resulting soft assignment matrix $A \in \mathbb{R}^{K \times H \times W}$ enables differentiable cluster-wise operations while allowing pixels at depth boundaries to smoothly transition between clusters.

\subsection{Flash Simulation Module}
\label{ssec:flash}

The flash simulation module $\F$ generates a virtual flash image using cluster-wise Gaussian illumination. Dark regions receive stronger enhancement while bright areas are preserved.
For each cluster $k$, we compute the weighted mean intensity $\bar{I}_k$ and maximum channel response $M_k$:
\begin{align}
    \bar{I}_k &= \frac{\sum_{i,j} A_k(i,j) \cdot \text{mean}(I_{low}(i,j))}{\sum_{i,j} A_k(i,j) + \epsilon}, \\
    M_k &= \frac{\sum_{i,j} A_k(i,j) \cdot \max_c(I_{low}^c(i,j))}{\sum_{i,j} A_k(i,j) + \epsilon},
\end{align}
where $\epsilon=10^{-8}$ ensures numerical stability and $c$ indexes the RGB channels.
The flash intensity for cluster $k$ is:
\begin{equation}
    \phi_k = \alpha \cdot (1 - \bar{I}_k) + \beta \cdot M_k + \gamma \cdot \mathcal{N}(0, 1),
\end{equation}
where $\alpha=1.5$, $\beta=0.3$, and $\gamma=0.1$ are parameters controlling base intensity, highlight preservation, and spatial variation respectively. The Gaussian noise term $\mathcal{N}(0,1)$ introduces natural-looking flash variation during training.
The final flashed image is obtained by accumulating cluster-weighted flash contributions:
\begin{equation}
    I_{flash} = \text{clamp}\left(I_{low} + \sum_{k=1}^{K} \phi_k \cdot A_k, 0, 1\right).
\end{equation}

\subsection{Flash Feature Encoder}
\label{ssec:flash_encoder}

To extract rich representations from the flash-simulated image, we employ a dedicated encoder network $\E$ with the same architectural structure as the depth estimation encoder, ensuring consistent feature dimensions for subsequent fusion:
\begin{equation}
    \{F^{(l)}_f\}_{l=1}^{5} = \E(I_{flash}),
\end{equation}
where $F^{(l)}_f$ represents flash features at scale level $l$.

\subsection{Main Enhancement Network}
\label{ssec:main_net}

The core of our architecture is an encoder-decoder structure augmented with Efficient Fusion Blocks (EFB) that integrate depth and flash features through multi-head attention.

\subsubsection{Efficient Fusion Block}

Direct application of self-attention to high-resolution feature maps has computational complexity $O((HW)^2)$. We propose an Efficient Fusion Block that reduces complexity to $O(P^2)$ where $P=8$ is a fixed pooling size. Given main encoder features $F_m^{(l)}$, depth features $F_d^{(l)}$, and flash features $F_f^{(l)}$ at level $l$, the EFB first applies adaptive average pooling: 
\begin{equation}
    \hat{F}_m = \text{AdaptivePool}_{P \times P}(F_m^{(l)}),
\end{equation}
The pooled features are flattened to $\hat{F}_m \in \mathbb{R}^{B \times P^2 \times C_l}$ for attention computation. Self-attention captures intra-feature dependencies as follows:
\begin{equation}
    F_{self} = \text{LayerNorm}(\hat{F}_m) + \text{MHA}(\hat{F}_m, \hat{F}_m, \hat{F}_m),
\end{equation}
where MHA denotes multi-head attention with 4 heads and 10\% dropout.

Cross-attention then integrates auxiliary depth and flash information. The concatenated auxiliary features are projected to match the main feature dimension:
\begin{equation}
    F_{aux} = \text{Conv}_{1 \times 1}([\hat{F}_d; \hat{F}_f]),
\end{equation}
\begin{equation}
    F_{cross} = \text{LayerNorm}(F_{self}) + \text{MHA}(F_{self}, F_{aux}, F_{aux}).
\end{equation}
The attended features are upsampled back to the original spatial resolution and refined through a convolutional feed-forward network:
\begin{equation}
    F_{out} = F_m^{(l)} + \text{FFN}(F_m^{(l)} + \gamma \cdot \text{Upsample}(F_{cross})),
\end{equation}
where $\gamma=0.1$ is a scaling parameter, and the FFN consists of $1 \times 1$ convolution, BatchNorm, GELU activation, and $3 \times 3$ convolution.

\subsubsection{Encoder-Decoder Architecture}

The main network follows an encoder–decoder design with residual enhancement.
The encoder has five levels with channels $[C,2C,4C,8C,16C]$ where $C=32$. Each level performs max-pooling and double convolution, followed by an EFB that fuses depth and flash features to $E^{(l)}$. The decoder uses bilinear upsampling with skip connections from fused encoder features and double-convolution refinement, producing $D^{(l)}$. Finally, a residual formulation predicts an enhancement map that is added to the input image, allowing the model to focus on learning only the residual details.
\begin{equation}
    E^{(l)} = \text{EFB}(\text{DoubleConv}(\text{MaxPool}(E^{(l-1)})), F_d^{(l)}, F_f^{(l)}),
\end{equation}
\begin{equation}
D^{(l)} = \text{DoubleConv}\left([D^{(l+1)}_{up}, E^{(l)}]\right),
\end{equation}
\begin{equation}
I_{\text{enh}} = \operatorname{clamp}\left(
\operatorname{Conv}_{1\times1}\!\left(D^{(1)}\right) + I_{\text{low}},
0, 1
\right).
\end{equation}

\section{Loss Functions}
\label{sec:loss}

Our training employs a multi-component loss as follows:
\begin{align}
\mathcal{L}_\text{total} &= \lambda_d \mathcal{L}_\text{depth} + \lambda_r \mathcal{L}_\text{recon} + \lambda_p \mathcal{L}_\text{perc} \notag \\
&\quad + \lambda_s \mathcal{L}_\text{ssim} + \lambda_c \mathcal{L}_\text{color} + \lambda_e \mathcal{L}_\text{edge} ,
\end{align}
where $\mathcal{L}_\text{depth}$, $\mathcal{L}_\text{recon}$, $\mathcal{L}_\text{perc}$, $\mathcal{L}_\text{ssim}$, $\mathcal{L}_\text{color}$, and $\mathcal{L}_\text{edge}$ represent depth supervision, reconstruction, perceptual, SSIM, color constancy, and edge preservation losses, respectively.

\subsection{Depth Consistency Loss}
\label{ssec:depth_loss}

To supervise the depth estimation network, we employ an L1 loss between the predicted depth map and the pseudo ground-truth obtained from the pretrained MiDaS~\cite{birkl2023midas} model on the well-lit image:
\begin{equation}
    \L_{depth} = \|D_{pred} -  \text{MiDaS}(I_{high})\|_1.
\end{equation}

\subsection{Reconstruction Loss}
\label{ssec:recon_loss}

The primary reconstruction objective is formulated as pixel-wise L1 loss between the enhanced output and ground truth:
\begin{equation}
    \L_{recon} = \|I_{enh} - I_{high}\|_1.
\end{equation}

\subsection{Perceptual Loss}
\label{ssec:perc_loss}

To capture high-level semantic similarity and encourage perceptually pleasing results, we employ a perceptual loss $L_{perc}$ based on VGG-19~\cite{simonyan2014very} features. Let $\Phi^{(j)}$ denote the $j$-th layer feature extractor of a pretrained VGG-19 network. The loss:
\begin{equation}
    \L_{perc} = \sum_{j \in \mathcal{J}} \|\Phi^{(j)}(I_{enh}) - \Phi^{(j)}(I_{high})\|_1,
\end{equation}
where $\mathcal{J} = \{3, 8, 17, 26, 35\}$ corresponds to relu1\_2, relu2\_2, relu3\_4, relu4\_4, and relu5\_4 layers. Input images are normalized using ImageNet statistics before feature extraction.

\subsection{Structural Similarity Loss}
\label{ssec:ssim_loss}

To preserve structural consistency, we incorporate SSIM loss:
\begin{equation}
    \L_{ssim} = 1 - \text{SSIM}(I_{enh}, I_{high}),
\end{equation}
where SSIM is computed using a Gaussian window of size $11 \times 11$ with $\sigma=1.5$:
\begin{equation}
    \text{SSIM}(x, y) = \frac{(2\mu_x\mu_y + C_1)(2\sigma_{xy} + C_2)}{(\mu_x^2 + \mu_y^2 + C_1)(\sigma_x^2 + \sigma_y^2 + C_2)}
\end{equation}
with stability constants $C_1 = 0.01^2$ and $C_2 = 0.03^2$.

\subsection{Color Consistency Loss}
\label{ssec:color_loss}

Low-light enhancement often suffers from color shifts. To mitigate this, we introduce a color consistency loss computed in the perceptually uniform CIELAB color space:
\begin{equation}
    \L_{color} = \|\text{RGB2LAB}(I_{enh}) - \text{RGB2LAB}(I_{high})\|_1.
\end{equation}
The RGB to LAB conversion involves linearization, transformation to XYZ color space, and subsequent conversion to LAB coordinates, ensuring that color differences are perceptually meaningful.

\subsection{Edge Preservation Loss}
\label{ssec:edge_loss}

We employ an edge preservation loss using Sobel gradient operators~\cite{gonzalez2009digital} to preserve edge structures and fine details:
\begin{equation}
    \L_{edge} = \|\nabla I_{enh} - \nabla I_{high}\|_1,
\end{equation}
where $\nabla I = \sqrt{(I * S_x)^2 + (I * S_y)^2 + \epsilon}$ computes the gradient magnitude using horizontal and vertical Sobel kernels $S_x$ and $S_y$:

\section{Experiments}
\label{sec:exp}

\subsection{Experimental Settings}
For evaluation of our LUMEN method, we conducted experiments on 3 widely used datasets with paired low-light and normal-light images: LOL-v1~\cite{wei2018deep}, LOL-v2 Real~\cite{yang2021sparse}, and LOL-v2 Synthetic~\cite{yang2021sparse}.
The LOL-v1 dataset comprises 485 pairs of low-light and normal-light images for training and 15 pairs for testing. Each image has a spatial resolution of 600 × 400 pixels, and all samples were captured under real-world illumination conditions. The LOL-v2 Real subset contains 689 training pairs and 100 testing pairs of real low-light and corresponding normal-light images. In contrast, the LOL-v2 Syn subset consists of 900 training pairs and 100 testing pairs, where the low-light images are synthetically generated from normal-light counterparts to simulate diverse lighting degradations.

Our proposed network is trained using the AdamW optimizer with $\beta_1=0.9$, $\beta_2=0.999$, and a weight decay of $10^{-4}$. The initial learning rate is set to $10^{-4}$ and adjusted using a Cosine Annealing scheduler down to a minimum of $10^{-6}$ over 100 epochs. The batch size is set to 8. For training, we resize all input images to 128$\times$128 pixels.
The weights in (16) are set to $\lambda_{r}=1.0$, $\lambda_{s}=0.5$, $\lambda_{p}=0.1$, $\lambda_{d}=0.5$, $\lambda_{c}=0.3$, and $\lambda_{e}=0.2$.

\subsection{Evaluation Metrics}
To quantitatively evaluate the performance of our method, we employ four standard metrics: peak signal-to-noise ratio (PSNR), structural similarity index measure (SSIM)~\cite{wang2004image}, learned perceptual image patch similarity (LPIPS)~\cite{zhang2018unreasonable}, and mean absolute error (MAE). 
Higher values for PSNR and SSIM indicate better performance, while lower values for LPIPS and MAE denote superior image restoration quality.

\subsection{Quantitative Comparison}
To comprehensively assess the performance of the proposed LUMEN framework, we compare it against a curated set of representative low-light image enhancement (LLIE) methods including LIME~\cite{guo2016lime}, RetinexNet~\cite{wei2018deep}, KinD~\cite{zhang2019kindling}, RUAS~\cite{liu2021retinex}, Zero-DCE~\cite{guo2020zero}, EnlightenGAN~\cite{jiang2021enlightengan}, CUE~\cite{zheng2023empowering}, CRetinex~\cite{xu2024cretinex} and DiffUIR~\cite{zheng2024selective}.




\begin{figure*}[!t]
\centering
\scriptsize
\setlength{\tabcolsep}{1pt}
\begin{tabular}{ccccc}
Input & RetinexNet & KinD & LUMEN (ours) & Ground Truth \\[0.1em]
\includegraphics[width=0.19\textwidth]{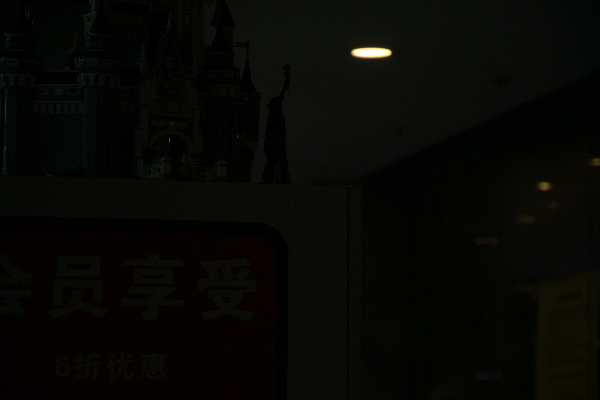} &
\includegraphics[width=0.19\textwidth]{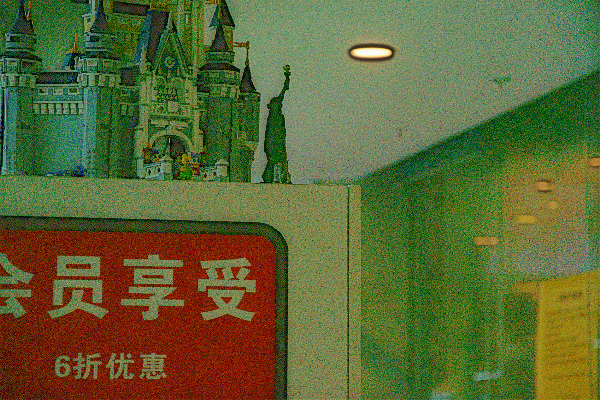} &
\includegraphics[width=0.19\textwidth]{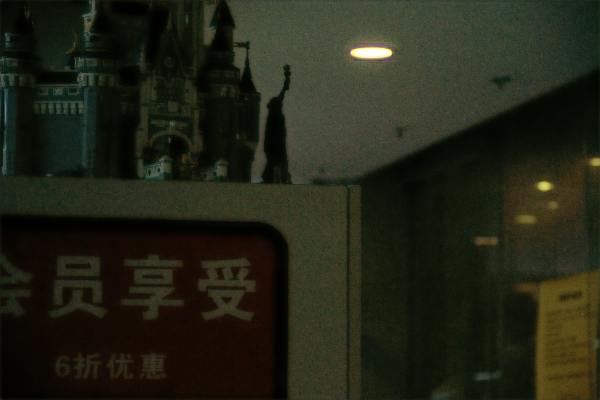} &
\includegraphics[width=0.19\textwidth]{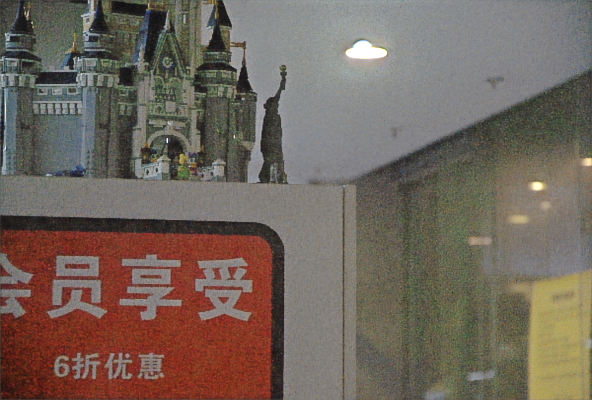} &
\includegraphics[width=0.19\textwidth]{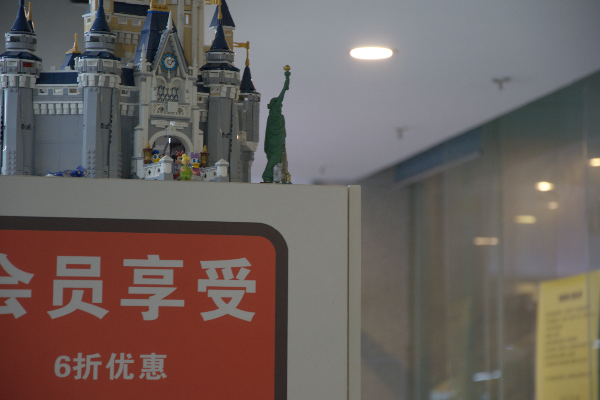} \\[0.2em]

\includegraphics[width=0.19\textwidth]{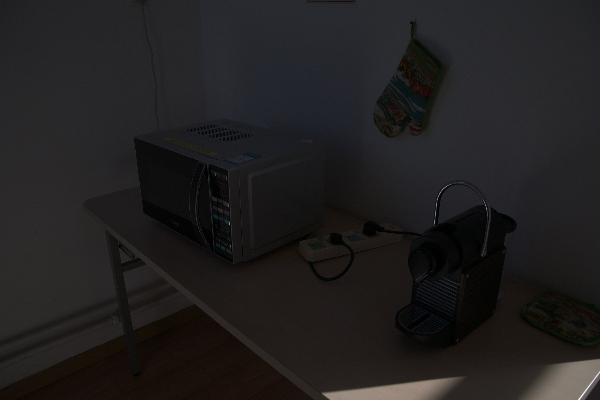} &
\includegraphics[width=0.19\textwidth]{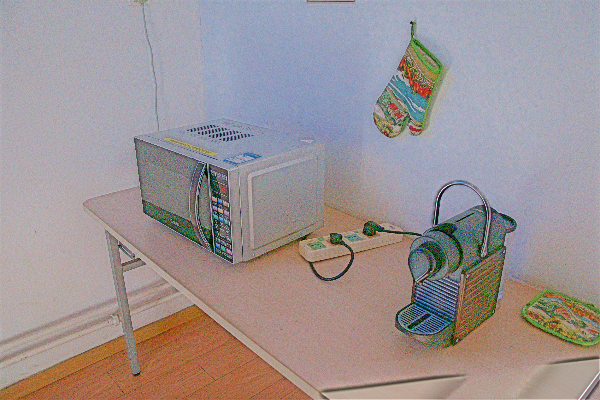} &
\includegraphics[width=0.19\textwidth]{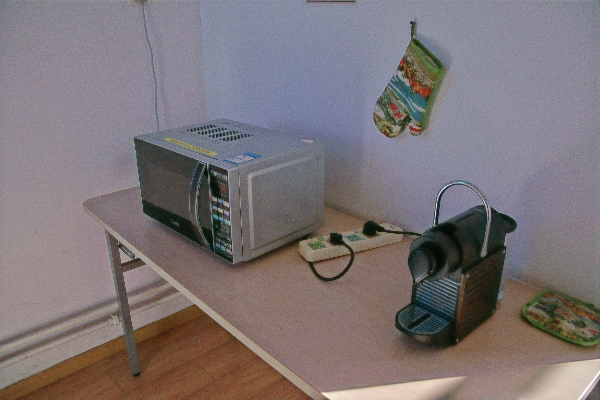} &
\includegraphics[width=0.19\textwidth]{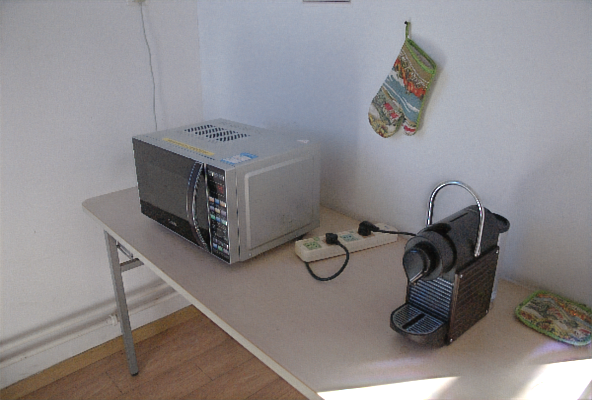} &
\includegraphics[width=0.19\textwidth]{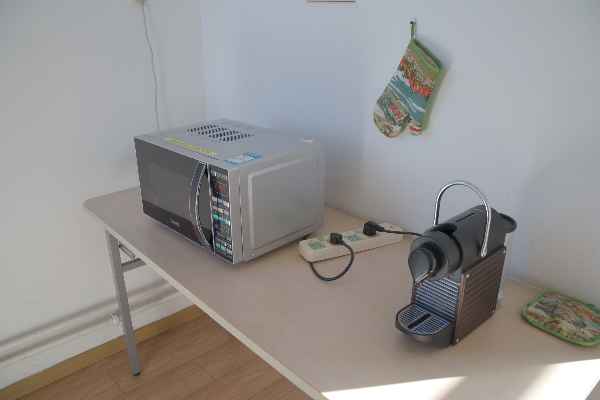} \\[0.2em]

\includegraphics[width=0.19\textwidth]{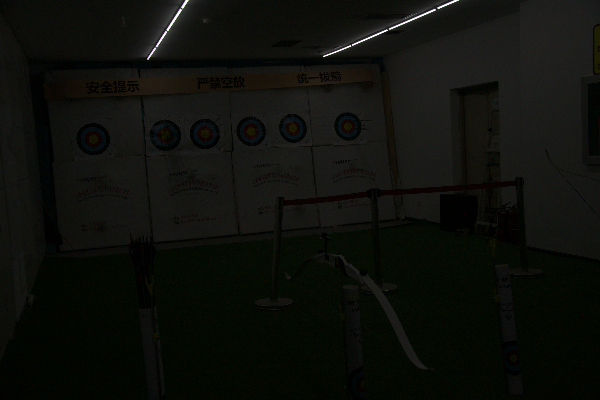} &
\includegraphics[width=0.19\textwidth]{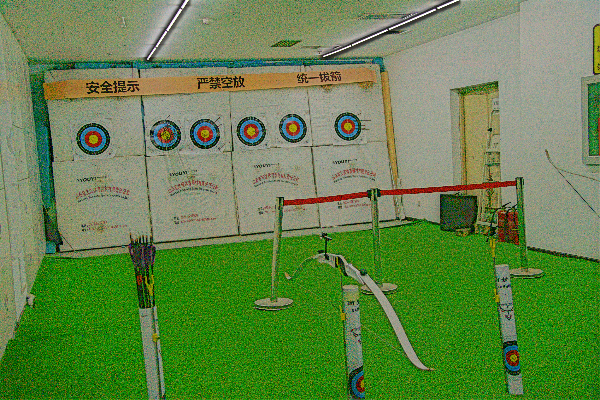} &
\includegraphics[width=0.19\textwidth]{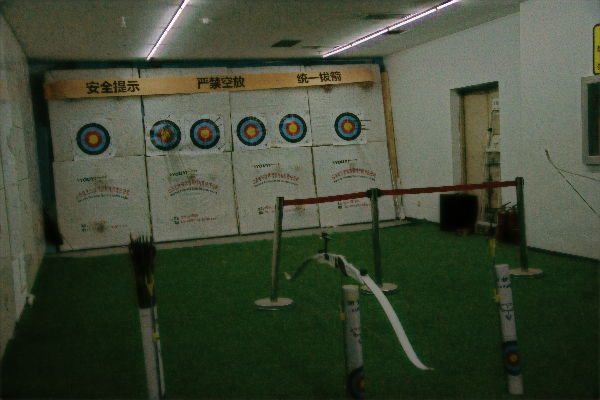} &
\includegraphics[width=0.19\textwidth]{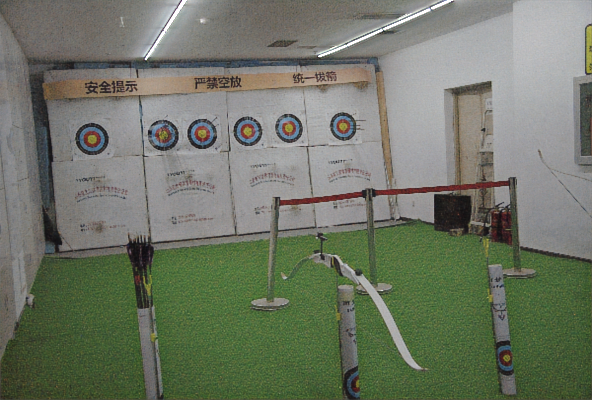} &
\includegraphics[width=0.19\textwidth]{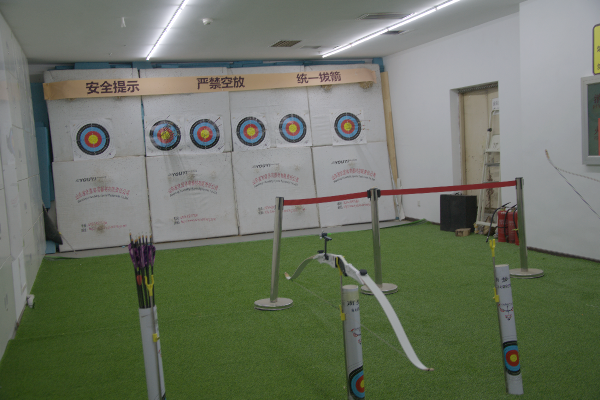}
\end{tabular}
\caption{Visual comparison of low-light image enhancement results on representative samples from the LOL dataset. }
\label{fig:comparison_visual}
\end{figure*}

\begin{table*}[!t]
\centering
\caption{Quantitative Comparison of LUMEN with Existing LLIE (Low-Light Image Enhancement) Methods on LOLv1, LOLv2 Real, and LOLv2 Syn datasets.}
\label{tab:LUMEN_LoL_full}
\renewcommand{\arraystretch}{1.25}
\setlength{\tabcolsep}{4pt}
\footnotesize
\begin{tabular}{l|l|cccc|cccc|cccc}
\hline
\multirow{2}{*}{\textbf{Methods}} 
& \multirow{2}{*}{\textbf{Venue}} 
& \multicolumn{4}{c|}{\textbf{LOLv1 Dataset}} 
& \multicolumn{4}{c|}{\textbf{LOLv2 Real Dataset}} 
& \multicolumn{4}{c}{\textbf{LOLv2 Syn Dataset}} \\
\cline{3-14}
& 
& PSNR$\uparrow$ & SSIM$\uparrow$ & LPIPS$\downarrow$ & MAE$\downarrow$
& PSNR$\uparrow$ & SSIM$\uparrow$ & LPIPS$\downarrow$ & MAE$\downarrow$
& PSNR$\uparrow$ & SSIM$\uparrow$ & LPIPS$\downarrow$ & MAE$\downarrow$ \\
\hline
LIME~\cite{guo2016lime} & TIP'17 
& 16.554 & 0.429 & 0.405 & 0.123
& 15.105 & 0.402 & 0.426 & 0.145
& 16.570 & 0.740 & 0.210 & 0.121 \\

RetinexNet~\cite{wei2018deep} & BMVC'18 
& 17.558 & 0.651 & 0.379 & 0.117
& 16.097 & 0.401 & 0.543 & 0.131
& 17.137 & 0.762 & 0.255 & 0.117 \\

KinD~\cite{zhang2019kindling} & ACMMM'19 
& 17.648 & 0.775 & 0.175 & 0.123
& 16.828 & 0.759 & 0.224 & 0.129
& 18.930 & 0.809 & 0.174 & 0.107 \\

RUAS~\cite{liu2021retinex} & CVPR'21 
& 16.405 & 0.500 & 0.270 & 0.153
& 15.326 & 0.488 & 0.310 & 0.162
& 13.404 & 0.645 & 0.364 & 0.196 \\

Zero-DCE~\cite{guo2020zero} & CVPR'20 
& 19.524 & 0.703 & 0.330 & 0.106
& 18.059 & 0.574 & 0.313 & 0.131
& 17.756 & 0.816 & 0.168 & 0.124 \\

EnlightenGAN~\cite{jiang2021enlightengan} & TIP'21 
& 17.483 & 0.651 & 0.322 & 0.135
& 18.640 & 0.676 & 0.309 & 0.132
& 16.573 & 0.775 & 0.212 & 0.133 \\

CUE~\cite{zheng2023empowering} & ICCV'23 
& 21.670 & 0.775 & 0.224 & 0.079
& 18.053 & 0.753 & 0.347 & 0.129
& \underline{22.209} & \underline{0.877} & \underline{0.117} & \underline{0.072} \\

CRetinex~\cite{xu2024cretinex} & IJCV'24 
& 19.866 & 0.807 & 0.196 & 0.096
& 18.265 & 0.789 & 0.285 & 0.122
& 20.366 & 0.876 & 0.125 & 0.086 \\

DiffUIR~\cite{zheng2024selective} & CVPR'24 
& \underline{22.407} & \underline{0.834} & \underline{0.157} & \underline{0.073}
& \textbf{19.710} & \underline{0.825} & \underline{0.211} & \textbf{0.105}
& 19.611 & 0.863 & 0.160 & 0.103 \\

\hline
\textbf{LUMEN} & \textbf{Ours} 
& \textbf{22.43} & \textbf{0.8846} & \textbf{0.0828} & \textbf{0.0700}
& \underline{19.27} & \textbf{0.8785} & \textbf{0.0852} & \underline{0.1081}
& \textbf{22.99} & \textbf{0.9014} & \textbf{0.0709} & \textbf{0.0679} \\
\hline
\end{tabular}
\end{table*}

\subsection{Qualitative Analysis}
To demonstrate the effectiveness of the proposed approach, we present a qualitative comparison against two representative state-of-the-art low-light enhancement methods, RetinexNet~\cite{wei2018deep} and KinD~\cite{zhang2019kindling}, as illustrated in Figure~\ref{fig:comparison_visual}. As observed, RetinexNet~\cite{wei2018deep} frequently amplifies sensor noise and introduces noticeable color distortions, which manifest as unnatural greenish artifacts on the floor in row~3 and pronounced graininess in the castle scene shown in row~1. While KinD~\cite{zhang2019kindling} preserves some structure, it often degrades contrast and fails to fully recover luminance in the darkest regions, resulting in undersaturated images. In contrast, our proposed method, LUMEN, consistently achieves superior perceptual quality by carefully balancing illumination enhancement with effective noise suppression. The method produces more natural color reproduction, improved local contrast, and better detail recovery. Furthermore, LUMEN maintains consistent performance across diverse scenes, demonstrating its robustness in handling varying illumination conditions.

\section{Conclusion}
\label{sec:conclusion}
In this work, we proposed a novel deep learning framework for low-light image enhancement that effectively integrates geometric priors with adaptive feature fusion. The core idea was to leverage monocular depth estimation to generate a simulated ``flash'' guide, followed by extracting and fusing these multi-scale structural cues into a main restoration network. The depth and flash modules adaptively revealed hidden details in dark regions, while the attention-based fusion blocks effectively integrated these features to recover structural integrity, suppress noise, and correct color deviations. Experimental results confirmed that our model successfully balances global illumination enhancement with local detail preservation, outperforming existing state-of-the-art approaches.

\bibliographystyle{IEEEbib}
\bibliography{ref}

\end{document}